\documentclass[%
 reprint,
 prl,
 amsmath,amssymb,
 aps,
]{revtex4-1}
\usepackage{mathrsfs}
\usepackage{color}
\usepackage{ulem}
\usepackage{graphicx}
\usepackage{dcolumn}
\usepackage{bm}

\def\vec#1{\boldsymbol #1}

\begin{document}

\preprint{APS/123-QED}

\title{Gap opening mechanism for correlated Dirac electrons 
in organic compounds $\alpha$-(BEDT-TTF)$_2$I$_3$ and $\alpha$-(BEDT-TSeF)$_2$I$_3$}

\author{Daigo Ohki$^1$}
\author{Kazuyoshi Yoshimi$^2$}
\author{Akito Kobayashi$^1$}
\author{Takahiro Misawa$^3$}
 \affiliation{$^1$Department of Physics, Nagoya University, Furo-cho, Chikusa-ku, Nagoya, 464-8602 Japan\\
  $^2$Institute for Solid State Physics, University of Tokyo, Chiba 277-8581, Japan\\
  $^3$Beijing Academy of Quantum Information Sciences, Beijing 100193, China
  }




\date{\today}

\begin{abstract}
To determine how electron correlations open a gap 
in two-dimensional massless Dirac electrons in the
organic compounds $\alpha$-(BEDT-TTF)$_2$I$_3$~[$\alpha$-(ET)$_2$I$_3$] 
and $\alpha$-(BEDT-TSeF)$_2$I$_3$~[$\alpha$-(BETS)$_2$I$_3$], 
we derive and analyze $ab$ $initio$ low-energy effective Hamiltonians 
for these two compounds. 
We find that the horizontal stripe charge ordering  
opens a gap in the massless Dirac electrons in $\alpha$-(ET)$_2$I$_3$, while an insulating phase without explicit 
symmetry breaking appears in $\alpha$-(BETS)$_2$I$_3$. 
We clarify that the combination of the anisotropic 
transfer integrals and the electron correlations
induces a dimensional reduction in the spin correlations, i.e., 
one-dimensional spin correlations develop in $\alpha$-(BETS)$_2$I$_3$.
We show that the one-dimensional spin correlations open a gap in the massless Dirac electrons.
Our finding paves the way for opening gaps for 
massless Dirac electrons 
using strong electron correlations.
\end{abstract}

\maketitle

{\it Introduction---}Dirac electrons in solids such as graphene \cite{Wallace, Novoselov}, 
bismuth \cite{Wolff, FukuyamaKubo}, 
and several organic conductors \cite{Kajita1992, Tajima2000, Kobayashi2004, Katayama2006, Kobayashi2007, Goerbig2008, Kajita2014, Tajima2006} exhibit
many intriguing physical properties such as quantum conduction 
associated with universal conductivity \cite{Shon}, 
large diamagnetism \cite{FukuyamaKubo}, 
and anomalous electron 
correlation effects \cite{TanakaOgata, Ishikawa, HirataNat2016, Matsuno2017, Matsuno2018}.
In particular, there has been much interest in opening gaps for massless 
Dirac electrons, since gap opening with band inversion 
can produce the topological insulators~\cite{KaneMele,FuKaneMele_PRL2007}. 
Even though the insulating phases are topologically trivial,
massive Dirac electrons in solids are expected to be 
useful for device applications because of their high mobility~\cite{Duplock_PRL2004,Balog_2010Nmat}. 
Electronic correlations, which are always present in solids, 
are expected to play an important role in gap opening for massless Dirac electrons.
As a canonical model for studying how electron correlations can open gaps for massless Dirac electrons, 
the Hubbard model on a honeycomb lattice has been studied~\cite{Meng_Nature2010,Sorella_SR2012}.
In the simple Hubbard model, it has been shown that the antiferromagnetic order opens a gap for massless Dirac electrons~\cite{Sorella_SR2012}.

The organic compounds
$\alpha$-(BEDT-TTF)$_2$I$_3$~[BEDT-TTF=bis(ethylenedithio)tetrathiafulvalene](ET) and $\alpha$-(BEDT-TSeF)$_2$I$_3$ [BEDT-TSeF=bis(ethylenedithio)tetraselenafulvalene](BETS)
offer an ideal platform for studying correlated Dirac electrons.
It has been noted that massless Dirac electrons appear around the Fermi energy in these compounds, owing to accidental degeneracy in the momentum space~\cite{Kajita1992, Tajima2000, Kobayashi2004, Katayama2006, Kobayashi2007, Goerbig2008, Kajita2014, Tajima2006, KitouSawaTsumuraya, TsumurayaSuzumura, SuzumuraTsumuraya}.
Both $\alpha$-(ET)$_2$I$_3$ and $\alpha$-(BETS)$_2$I$_3$ have four ET and BETS molecules in a unit cell and 
inversion symmetry exists at high temperatures in the two-dimensional (2D) conduction plane composed of ET and BETS molecules.
Because of their similar crystal structures, the band structures
of both compounds are basically the same~\cite{KitouSawaTsumuraya}.
However, they have rather different insulating phases at low temperatures and this difference can be induced by strong electron correlations. 
As we show later, both $\alpha$ compounds
are located in the strongly correlated region since
the on-site Coulomb $U$ is larger than the bandwidth $W$ ($U/W>1$). 

In $\alpha$-(ET)$_2$I$_3$,
it has been reported that as the temperature is reduced, the horizontal stripe charge ordering (HCO) associated with inversion symmetry breaking
induces a gap for massless Dirac electrons~\cite{Seo2000, Takahashi, Kakiuchi}. 
Electronic correlations play important roles in both massless Dirac electrons and massive Dirac electrons.
In the massless Dirac electron phase, theoretical studies and nuclear magnetic resonance (NMR) experiments under an in-plane magnetic field have shown evidence of velocity renormalization, reshaping of the Dirac cone, and weak ferrimagnetic spin polarization caused by Coulomb interactions \cite{HirataScience, Matsuno2018, Ohki2020, HirataNat2016, Matsuno2017}.
In the HCO insulator phase, it has been suggested that anisotropy of nearest-neighbor Coulomb interactions in the 2D plane is the origin of the HCO phase transition of $\alpha$-(ET)$_2$I$_3$~\cite{Seo2000}.
In the vicinity of the phase transition, $\alpha$-(ET)$_2$I$_3$ exhibits 
anomalous properties for the spin gap~\cite{TanakaOgata, Ishikawa} and 
transport phenomena~\cite{Beyer, Liu, Ohki2019}.

$\alpha$-(BETS)$_2$I$_3$ has a distinctly different insulating state to $\alpha$-(ET)$_2$I$_3$.
It has been reported that the direct-current resistivity becomes almost constant, related to the universal conductivity, at $T>50$ K and sharply increases at $T<50$~K~\cite{Inokuchi, Kawasugi, TajimaPriv}.
This result suggests that a charge gap opens below $50$~K.
However, no signatures of spatial inversion symmetry breaking or changes in bond length between nearest-neighbor BETS molecules have been found~\cite{KitouSawaTsumuraya}.
These experimental results indicate that the gap opening mechanism in $\alpha$-(BETS)$_2$I$_3$ cannot be attributed to simple charge and/or magnetic ordering.
The $ab$ $initio$ band calculations suggest that the gap can be opened by spin-orbit coupling (SOC) in  $\alpha$-(BETS)$_2$I$_3$.
However, the gap estimated by SOC ($\sim$ 2~meV)~\cite{Winter, TsumurayaSuzumura, SuzumuraTsumuraya} 
is too small to account for the insulating behavior below $50$~K.
Therefore, the mechanism of gap opening has not yet been fully clarified. 

In this Letter, to determine the origin of the differences 
in the gap opening mechanisms in $\alpha$-(ET)$_2$I$_3$ and $\alpha$-(BETS)$_2$I$_3$, 
{we employ an $ab$ $initio$ method for correlate electron systems~\cite{Imada_JPSJ2010},
which succeeds in reproducing the electronic structures of several molecular solids~\cite{Shinaoka_JPSJ2012,Misawa_PRR2020,Yoshimi_PRR2021,Ido_npj2022}.}
{In the method,}
we first derive $ab$ $initio$ low-energy effective Hamiltonians.
Then, we solve the effective Hamiltonians using {accurate low-energy solvers such as}
the many-variable variational Monte Carlo method (mVMC)~\cite{mVMC}.
Based on this, it is found that a HCO insulating state appears in $\alpha$-(ET)$_2$I$_3$, which is consistent with experiments and previous studies. 
However, in $\alpha$-(BETS)$_2$I$_3$, we find that an insulating state without any explicit symmetry breaking is realized.
Because of the frustration in the inter-chain magnetic interactions, we find that dimensional reduction of the spin correlations occurs, i.e., one-dimensional spin correlations develop in a certain chain of $\alpha$-(BETS)$_2$I$_3$.
This result demonstrates that the one-dimensional spin correlation is the main driver inducing
the gap in $\alpha$-(BETS)$_2$I$_3$, as in the one-dimensional Hubbard model~\cite{Lieb_PRL1968}.
Our calculation demonstrates that $\alpha$-(BETS)$_2$I$_3$ hosts massive Dirac electrons without symmetry breaking via dimensional reduction.

\begin{figure}
\begin{centering}
\includegraphics[width=80mm]{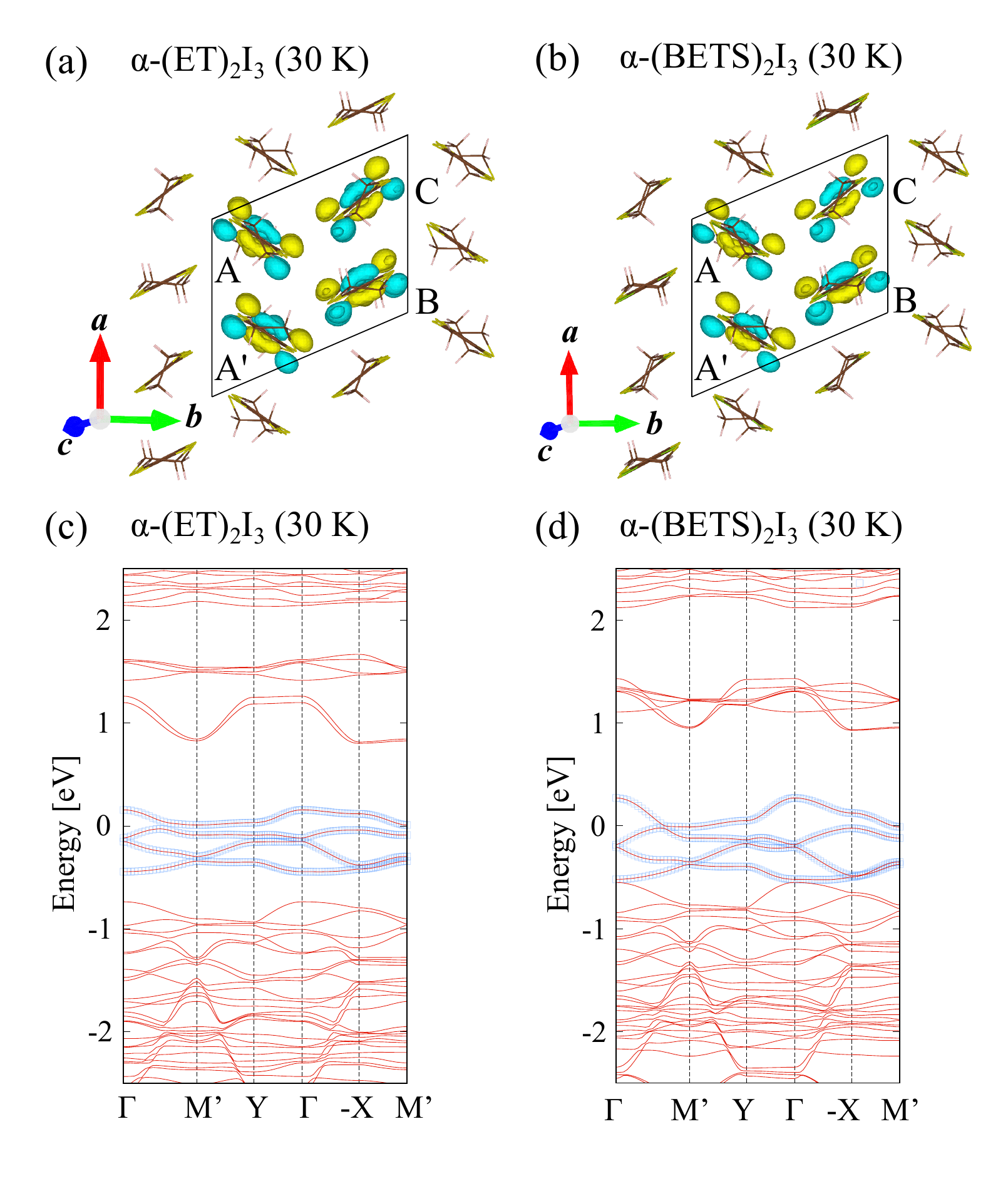}
\caption{(Color online) 
Crystal structures and real-space distribution of MLWFs for (a) $\alpha$-(ET)$_2$I$_3$ and (b) $\alpha$-(BETS)$_2$I$_3$ at 30 K drawn by VESTA \cite{Momma}.
Four ET(BETS) molecules (labeled by A, A$'$, B, and C sites) exist in the unit cell indicated by the black lines.
The A and A$'$ are crystallographically equivalent.
Energy band structures for (c) $\alpha$-(ET)$_2$I$_3$ and (d) $\alpha$-(BETS)$_2$I$_3$ at 30 K. The solid lines are obtained by the DFT calculations, while the squares are obtained from the MLFWs. Here, we define $\Gamma \equiv (0, 0, 0)$, M' $\equiv (-\pi, \pi, 0)$, Y$ \equiv (0, \pi, 0)$, X $\equiv (\pi, 0, 0)$. 
The bandwidth for the four bands of $\alpha$-(ET)$_2$I$_3$ is approximately $3/4$ times smaller than that of $\alpha$-(BETS)$_2$I$_3$.
}\label{Fig:QE_Wannier}
\end{centering}
\end{figure}

{\it Ab initio calculations---}We perform $ab$ $initio$ calculations to derive the effective Hamiltonians 
using the crystal structure data for $\alpha$-(ET)$_2$I$_3$ and $\alpha$-(BETS)$_2$I$_3$ at $T=30$~K~\cite{KitouSawaTsumuraya}.
Quantum ESPRESSO~\cite{Perdew,Giannozzi} with the SG15 optimized norm-conserving Vanderbilt pseudopotentials \cite{Schlipf} 
is used to obtain the global band structures by the density functional theory~(DFT) calculations~\cite{DFT}.
We construct maximally localized Wannier functions (MLWFs) using RESPACK~\cite{Nakamura}.
Figures \ref{Fig:QE_Wannier}(a) and (b) show the schematic crystal structure and 
real-space distribution of MLWFs for $\alpha$-(ET)$_2$I$_3$ and $\alpha$-(BETS)$_2$I$_3$ at $30$~K, respectively. 
Both $\alpha$-(ET)$_2$I$_3$ and $\alpha$-(BETS)$_2$I$_3$ have four BETS and ET molecules (sites) labeled A, A$'$, B, and C in the unit cell.  In $\alpha$-(BETS)$_2$I$_3$, the A and A$'$ sites are crystallographically equivalent due to inversion symmetry.
The calculation results for the energy bands obtained by the DFT calculations and MLFWs for $\alpha$-(ET)$_2$I$_3$ and $\alpha$-(BETS)$_2$I$_3$ at $30$~K are plotted as solid lines and symbols in Figs. \ref{Fig:QE_Wannier}(c) and (d), respectively.
The energy origin is set to be the Fermi energy.
We can see that the MLWFs reproduce the original band structures well.

Using the MLWFs, we evaluate the transfer integrals for these compounds and the screened Coulomb interactions using the constrained random phase approximation (cRPA).
The cutoff energy for the dielectric function is set at 5.0 Ry.
The obtained effective Hamiltonian is given by
\begin{align}
&H=\sum_{{\bm R},{\bm \delta}}\sum_{\alpha,\beta,\sigma}
(t^{(\bm \delta)}_{(\alpha,\beta)}c^{\dag}_{{\bm R},{\alpha},{\sigma}}c_{{\bm R}+{\bm \delta},\beta,{\sigma}}+{\rm H.c.}) \notag\\
&+\sum_{{\bm R},\alpha} U_\alpha n_{{\bm R},\alpha,\uparrow}n_{{\bm R},\alpha,\downarrow} \nonumber
+\sum_{{\bm R},{\bm \delta}}\sum_{\alpha,\beta}
V^{({\bm \delta})}_{(\alpha,\beta)}N_{{\bm R},\alpha} N_{{\bm R}+{\bm \delta},\beta} \label{Eq:Hamiltonian}
\end{align}
where ${\bm R}$ denotes the unit cell coordinate, 
and the 
{orbital} and spin indices are indicated by $\alpha$, $\beta$ (A, A', B, C)  and $\sigma$ (+1: $\uparrow$, -1:$\downarrow$), respectively. The transfer integrals from $(\beta,\sigma)$ to $(\alpha,\sigma)$ separated by ${\bm \delta}$ are represented by $t^{({\bm \delta})}_{(\alpha,\beta)}$.
The creation and annihilation operators are denoted by
$c^{\dag}_{{\bm R},{\alpha},{\sigma_1}}$ and $c_{{\bm R},{\alpha},{\sigma_1}}$, respectively.
The number operators are defined as $n_{{\bm R},\alpha,\sigma}=c^{\dag}_{{\bm R},{\alpha},{\sigma}}c_{{\bm R},{\alpha},{\sigma}}$ and $N_{{\bm R},\alpha}=n_{{\bm R},\alpha,\uparrow}+n_{{\bm R},\alpha,\downarrow}$.
To reflect the two dimensionality of the effective Hamiltonians,
we subtract a constant value $\Delta_{\rm DDF}$ from the on-site and off-site Coulomb interactions.
Following a previous study, we take $\Delta_{\rm DDF}=0.20$eV for both compounds~\cite{Nakamura_PRB2012}.
We confirm that the value of the constant shift does not change the result significantly. 

\begin{figure}
\begin{centering}
\includegraphics[width=85mm]{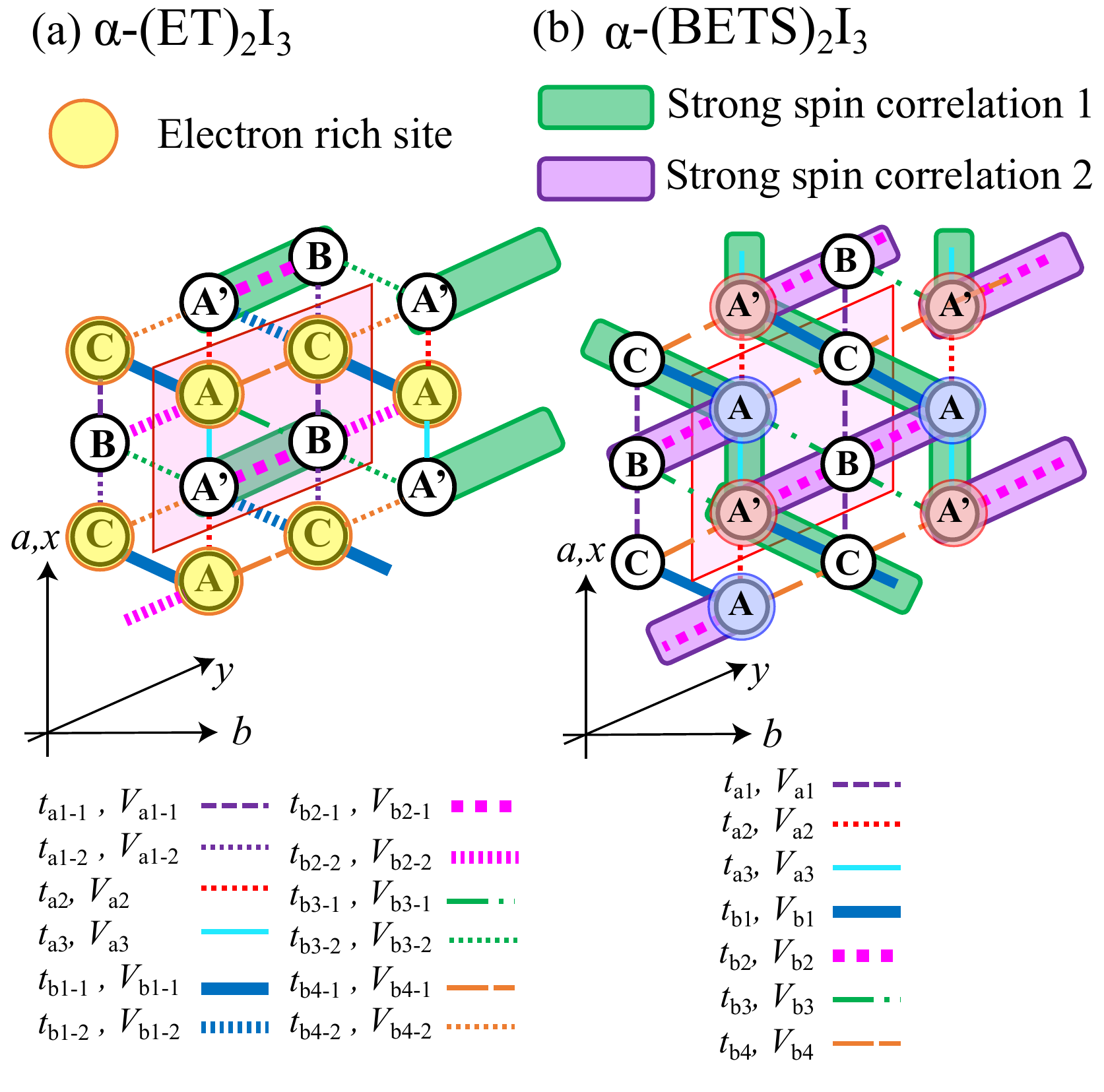}
\caption{(Color online) 
{Schematic diagrams of $\alpha$-type organic conductors
for (a) $\alpha$-(ET)$_2$I$_3$ and (b) $\alpha$-(BETS)$_2$I$_3$ at 30K.}
Transfer integrals and Coulomb interactions for the nearest-neighbor sites are also shown.
The shaded pink parallelogram shows a unit cell.
{We also show schematic} picture of the ground state 
obtained by mVMC for (a)~$\alpha$-(ET)$_2$I$_3$: horizontal stripe charge order (HCO) with spin dimer on strong transfer $t_{b2}$, and
(b)~$\alpha$-(BETS)$_2$I$_3$: one-dimensional antiferromagnetism (AF) correlations develop in A--A$^{\prime}$ chain.
Molecules surrounded by shaded purple and green rectangles indicate bonds with a strong 
spin singlet correlation, and molecules with a shaded yellow circle are electron-rich sites. 
}\label{Fig:mvmc_solutions}
\end{centering}
\end{figure}

{Figure~\ref{Fig:mvmc_solutions} shows schematic diagrams of the 
2D conduction plane of (a)~$\alpha$-(ET)$_2$I$_3$ and (b)~$\alpha$-(BETS)$_2$I$_3$,
showing the networks of transfer integrals and Coulomb interactions 
between the nearest-neighbor sites.}
We provide the values of the transfer integrals and the Coulomb interactions in the Supplemental materials~\cite{supplement} {and the raw data in the repository~\cite{data}}.
In both materials, the $b$-axis direction transfer integrals $t_{b1}$ and $t_{b2}$ 
are approximately 10 times larger than the others and make a strong transfer chain along the $b$-axis direction.
We note that in $\alpha$-(BETS)$_2$I$_3$, the strength of the transfer integral for $b1$ bond (A$'$-C, $t_{b1}=138.1$~meV) is comparable to that for $b2$ bond (A$'$-B, $t_{b2}=158.7$~meV). 
This indicates that the magnetic interactions between the A-A$'$ chain and the B-C chain are frustrated. 
This geometrical frustration induces a dimensional reduction in the spin correlations as we show later.
We can also see that the Coulomb interactions 
in $\alpha$-(ET)$_2$I$_3$ are around $1.25$ times 
larger than those in $\alpha$-(BETS)$_2$I$_3$. 


{\it mVMC analysis---}To investigate the ground states
of the effective Hamiltonians, we use the many-variable variational Monte Carlo (mVMC) method~\cite{mVMC}.
The trial wave function used in this study is given by
\begin{equation}
|\psi\rangle=\mathcal{P}_G\mathcal{P}_J\mathcal{L}_S|\phi_{\rm pair}\rangle,
\end{equation}
{where $\mathcal{L}_S$ represents the total spin projector
and we use the spin singlet projection for the ground states.} 
{The Gutzwiller factor $\mathcal{P}_G$ and the
Jastrow factor $\mathcal{P}_J$ are defined by
\begin{align}
&\mathcal{P}_G=\exp\Big[\sum_{i}g_{i}n_{i,\uparrow}n_{i,\downarrow}\Big],\\
&\mathcal{P}_J=\exp\Big[\frac{1}{2}\sum_{i\neq j}v_{ij}N_{i}N_{j}\Big],
\end{align}
where we denote the combination of the unit cell coordinate and the orbital index as $i=(\vec{R},\alpha)$.
}
The pair product part of the wave function $|\phi_{\rm pair}\rangle$ is defined as
\begin{equation}
|\phi_{\rm pair}\rangle=\left[\sum_{i,j}^{N_{\rm site}}f_{ij}c^{\dagger}_{i,\uparrow}c^{\dagger}_{j,\downarrow}\right]^{N_{\rm e}/2}|0\rangle,
\end{equation}
where $N_{\rm site}$ and $N_{\rm e}$ represent the total number of sites and electrons, respectively.
All variational parameters in the wavefunction are simultaneously 
optimized using the stochastic reconfiguration method~\cite{Sorella}.
We perform calculations for $L=4,6,8,10,12$ ($N_{\rm site} = 4\times L^2$)
with periodic boundary conditions. 
In the actual calculations, we impose a 2$\times$2 sublattice structure for the variational parameters.
We take hopping parameters up to $\vec{R}=(\pm2,\pm2)$ 
and Coulomb interactions up to the nearest-neighbor bonds shown in Fig.~\ref{Fig:mvmc_solutions}(a).
We also employ a particle--hole transformation to reduce the numerical cost.

{Figures~\ref{Fig:mvmc_solutions}(a) and \ref{Fig:mvmc_solutions}(b)} also show the schematic 
{charge configurations and spin correlations} in real space for the ground states for 
$\alpha$-(ET)$_2$I$_3$ and $\alpha$-(BETS)$_2$I$_3$.
In $\alpha$-(ET)$_2$I$_3$, the HCO insulator state is the ground state. 
The electron densities for $L=12$ at each site are
$\langle n_{\rm A}\rangle=1.58$, 
$\langle n_{\rm{A}'}\rangle=1.44$, 
$\langle n_{\rm B}\rangle=1.47$, and $\langle n_{\rm C}\rangle=1.51$.
Statistical errors in Monte Carlo sampling for the electron densities are on the order of $10^{-4}$.
We confirm that the system size dependence of local physical quantities such as electron density and spin correlation is small and, thus, in the following we show the result for $L=12$.
In the HCO state, the spin correlation for the {$b2{\rm-}1$} bond becomes large
$\langle {\bm S}_{\rm{A}'}\cdot{\bm S}_{\rm {B}}\rangle_{{b2{\rm-}1}}=-0.148(3)$, while
the spin correlation for the {$b3{\rm-}2$} bond becomes small, 
$\langle {\bm S}_{\rm B}\cdot{\bm S}_{\rm{A}'}\rangle_{{b3{\rm-}2}}=-0.016(1)$.
The parentheses denote the error--bars in the last digit. 
{Because of the HCO, ths spin correlations between charge rich sites become small.
For example, although the transfer integral of $b1{\rm-}1$ is comparable to that
of $b2{\rm-}1$ ($t_{b1{\rm-}1}=97.48$meV and $t_{b2{\rm-}1}=136.2$meV),
the spin correlation of $b1{\rm-}1$ bond is suppressed as
$\langle {\bm S}_{\rm{A}}\cdot{\bm S}_{\rm {C}}\rangle_{{b1{\rm-}1}}=-0.034(1)$.} 
These results indicate that the singlet dimer state associated 
with the emergence of HCO appears for the $b2{\rm-}1$ bond 
as shown in {Fig. \ref{Fig:mvmc_solutions}(a)}, which is consistent with the results of the NMR experiment and a previous theoretical study~\cite{Ishikawa, TanakaOgata}.
By analyzing the effective Hamiltonians for the 150 K structure, 
we find that the Coulomb interactions induce instability toward the HCO state, 
although lattice distortion is important for stabilizing the HCO state.
Details are shown in S.2 in Ref.~[\onlinecite{supplement}].

\begin{figure}[tb]
\begin{centering}
\includegraphics[width=85mm]{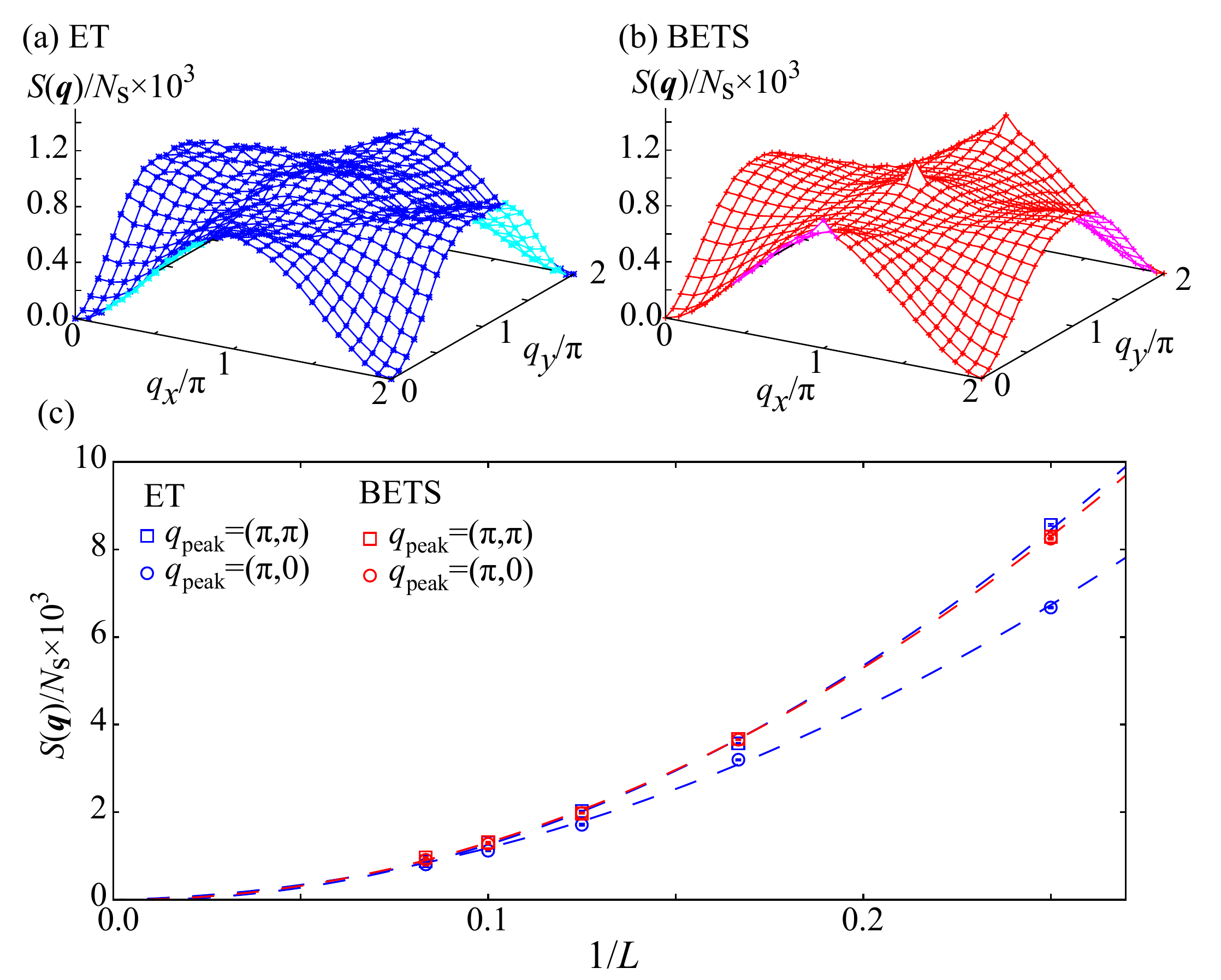}
\caption{(Color online) 
Spin structure factors for (a)$\alpha$-(ET)$_2$I$_3$ and (b)$\alpha$-(BETS)$_2$I$_3$.
We map the lattice structures into $2L\times2L$ square lattices.
The superposition of the $(\pi,0)$ and $(\pi,\pi)$ spin structures is consistent with the schematic images in Fig.~\ref{Fig:mvmc_solutions}.
(c)~System size dependence of peak values of spin structure factors.
The broken curves show the results of fitting using the function $a(1/L)+b(1/L)^2$.
}\label{Fig:Sq}
\end{centering}
\end{figure}

\begin{figure}[tb]
\begin{centering}
\includegraphics[width=85mm]{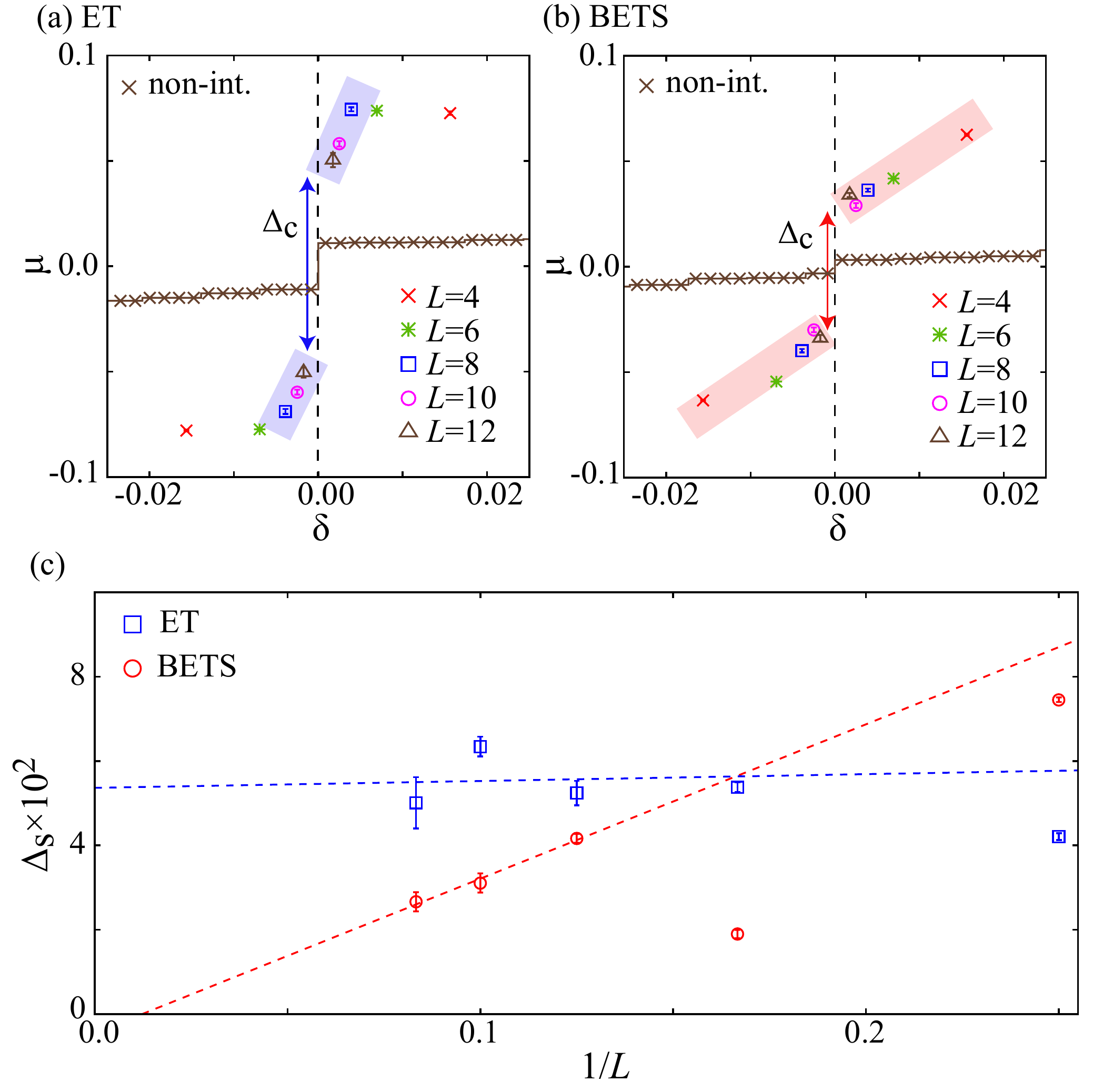}
\caption{(Color online) 
Doping dependence of chemical potential for 
(a)$\alpha$-(ET)$_2$I$_3$ and (b)$\alpha$-(BETS)$_2$I$_3$,
where $\mu_{0}=[\mu(N_{0}+1)-\mu(N_{0}-1)]/2$ and $N_{0}/N_{\rm s}=1.5$.
For $\alpha$-(ET)$_2$I$_3$ and $\alpha$-(BETS)$_2$I$_3$, 
the estimated charge gap is $\Delta_{\rm c}\sim0.1$eV and $\Delta_{\rm c}\sim0.07$eV.
For comparison, we plot the chemical potential 
for non-interacting systems for $L=12$ (brown crosses).
(c) Size dependence of the spin gap. We fit the
data for $L\geq8$ using the linear function $a+b(1/L)$
{to reduce the finite-size effects}.
}\label{Fig:mu}
\end{centering}
\end{figure}

For $\alpha$-(BETS)$_2$I$_3$, we cannot find any clear signature of the charge ordering.
The electron densities at each site are given by
$\langle n_{\rm A}\rangle=1.49$,
$\langle n_{\rm A'}\rangle=1.49$,
$\langle n_{\rm B}\rangle=1.50$,
and $\langle n_{\rm C}\rangle=1.52$.
Statistical errors in the Monte Carlo sampling are in order of $10^{-4}$.
$n_{\rm A}=n_{\rm A'}$ indicates that inversion symmetry is not broken.
We find that the spin correlations become strong for the $a2$, $b1$, and $b2$ bonds.
The spin correlations for these bonds are given by
$\langle {\bm S}_{\rm A}\cdot{\bm S}_{\rm A'}\rangle_{a3}=-0.0694(3)$, 
$\langle {\bm S}_{\rm A'}\cdot{\bm S}_{\rm C}\rangle_{b1}=-0.0735(6)$, 
and $\langle {\bm S}_{\rm A'}\cdot{\bm S}_{\rm B}\rangle_{b2}=-0.087(2)$.
These antiferromagnetic spin correlations 
are schematically shown in {Fig.~\ref{Fig:mvmc_solutions}(b)}.
This result indicates that the magnetic interactions
between A--A$'$ and B--C chains are frustrated.
Because of the inter-chain frustration,
long-range antiferromagnetic order is absent in $\alpha$-(BETS)$_2$I$_3$. 

Figures~\ref{Fig:Sq}(a) and (b) show the spin structure factors defined as
\begin{align}
S(\vec{q})=\frac{1}{N_{\rm s}}\sum_{i,j}\langle \vec{S}_{i}\cdot\vec{S}_{j} \rangle e^{i\vec{q}(\vec{r}_{i}-\vec{r}_{j})},
\end{align}
where we map the lattice structures to the $2L\times 2L$ square lattice 
(the directions of the $x$ and $y$ axes are shown in Fig.~\ref{Fig:mvmc_solutions}).
In the actual calculation, we limit the summation of one index to
within the unit cell to reduce the numerical cost.
For $\alpha$-(ET)$_{2}$I$_{3}$, we find no significant peaks
in the spin structure factors. This broad spin structure factor
is consistent with the one-dimensional spin dimer structures 
in the A$'$--B chain, as shown in {Fig.~\ref{Fig:mvmc_solutions}(a)}.

We find that peaks appear at $\vec{q}=(\pi,0)$ and $(\pi,\pi)$
in $\alpha$-(BETS)$_2$I$_3$. 
The superposition of the $(\pi,0)$ and $(\pi,\pi)$ order
indicates the emergence of the antiferromagnetic chain in the A--A$'$ chain.
Thus, the spin structure factor is consistent with the real space configuration 
in {Fig.~\ref{Fig:mvmc_solutions}(b)}.
However, the peak values become zero in the thermodynamic limit, as shown in 
Fig.~\ref{Fig:Sq}(c). This result indicates that 
the one-dimensionality of the spin correlations prohibits 
long-range magnetic order even at zero temperature.
Nevertheless, as we show below, the charge gap is finite
due to the one-dimensional spin correlations.

Here, we discuss the charge and spin gap in $\alpha$-(ET)$_{2}$I$_{3}$
and $\alpha$-(BETS)$_{2}$I$_{3}$.
In Figs.~\ref{Fig:mu}(a) and (b), we plot the chemical potential
$\mu(N+1)=[E(N+2)-E(N)]/2$ ($E(N)$ is the total energy for $N$-electrons systems)
as a function of the doping rate $\delta=N/N_{\rm s}-1.5$.
From this plot, we estimate the charge gap to be $\Delta_{\rm c}\sim 0.1$eV ($\Delta_{\rm c}\sim 0.07$eV) for $\alpha$-(ET)$_{2}$I$_{3}$ ($\alpha$-(BETS)$_{2}$I$_{3}$). The amplitude of the charge gap in  $\alpha$-(ET)$_{2}$I$_{3}$
is consistent with the experimental charge gap ($\Delta_{\rm c} \sim 0.07$eV) 
estimated from the optical conductivity~\cite{Clauss_PhysicaB2010}.
In $\alpha$-(ET)$_{2}$I$_{3}$, the existence of the charge gap is natural since 
the HCO and associated inversion symmetry breaking can open a gap for the massless Dirac electrons.
However, the charge gap in $\alpha$-(BETS)$_{2}$I$_{3}$ 
cannot be explained by simple symmetry breaking since there is no
clear signature of spin and charge ordering.
This result indicates that the one-dimensional antiferromagnetic correlations
developed in the A--A$'$ bonds induce the gap for massless Dirac electrons.
We note that the amplitude of the charge gap is sufficiently 
larger than that of the finite-size gap, which is about 0.01 eV.
This indicates that the finite charge gap obtained by the mVMC calculations
is not an artifact due to the finite system size.

Figure~\ref{Fig:mu}(c) shows the size dependence of the spin gap, defined as $\Delta_{\rm s}=E(S=1)-E(S=0)$.
Using the spin quantum number projection, we obtain the energy of the
triplet $(S=1)$ excited state.
Although the size dependence is not smooth, it is 
likely that the spin gap is finite in the thermodynamic limit
for $\alpha$-(ET)$_{2}$I$_{3}$. 
This is consistent with the existence of a spin dimer chain in A$^{\prime}$--B bonds.
The amplitude of the spin gap, $\Delta_{\rm s}\sim0.05$eV, is also consistent with the experimental result~\cite{Ishikawa}.
For $\alpha$-(BETS)$_{2}$I$_{3}$, the spin gap monotonically decreases except for $L=6$.
A size extrapolation using data for $L\geq8$ indicates that
the spin gap almost vanishes in the thermodynamic limit.
From the present calculation, although it is difficult
to accurately estimate the spin gap in the thermodynamic limit,
it is reasonable to conclude that the spin gap in $\alpha$-(BETS)$_{2}$I$_{3}$ is significantly
smaller than that in $\alpha$-(ET)$_{2}$I$_{3}$.

{\it Summary and Discussion---}In this study, to determine the origin of gap opening for massless Dirac electrons in $\alpha$-(ET)$_2$I$_3$ and $\alpha$-(BETS)$_2$I$_3$, 
we derive the low-energy effective Hamiltonians 
and solve them using the mVMC method~\cite{mVMC}. 
We find that the HCO insulator state appears in $\alpha$-(ET)$_2$I$_3$ 
while no clear symmetry breaking occurs in $\alpha$-(BETS)$_2$I$_3$. 
Nevertheless, we find that a charge gap opens in $\alpha$-(BETS)$_2$I$_3$
due to the development of one-dimensional spin correlations in the A--A$'$ chain.
We note that the recent observed increase in $1/(T_{1}T)$ of NMR below 20 K is 
consistent with the development of the one-dimensional spin correlations \cite{Fujiyama}.
We also note that weak but finite three dimensionality, 
which is not included in this study,
can induce long-range magnetic order at low temperatures
since the one-dimensional spin correlations are already developed in the conducting layer. 
Thus, the recently discovered antiferromagnetic order at low temperatures 
is consistent with our results~\cite{Konoike}.
Lastly, we consider the effects of spin--orbit coupling.
Although spin--orbit coupling alone cannot explain the
amplitude of the charge gap in $\alpha$-(BETS)$_2$I$_3$,
the combination of the Coulomb interactions and spin--orbit coupling is 
intriguing since it can enhance the SOC effectively and stabilize 
the quantum spin Hall insulating phase~\cite{Raghu, OhkiQSH} {or the three-dimensional
topological insulator~\cite{Nomoto_arxiv2022}.}
To examine such effects, it is necessary to derive and solve $ab$ $initio$ 
Hamiltonians with spin--orbit coupling.
This is an intriguing challenging issue but is left for future studies.

\begin{acknowledgements}
The authors would like to thank H. Sawa, T. Tsumuraya, and S. Kitou for their valuable comments.
We would like to express our gratitude to N. Tajima and Y. Kawasugi for informative 
discussions on the experimental aspects.
The computation in this work was performed using the 
facilities of the Supercomputer Center, Institute for Solid State Physics, University of Tokyo.
This work was supported by MEXT/JSPJ KAKENHI 
under grant numbers 21H01041, 19J20677, 19H01846, 15K05166 and 22K03526.
KY and TM were supported by Building of Consortia for the Development of Human Resources in Science and Technology, MEXT, Japan.
This work was also supported by the
National Natural Science Foundation of China (Grant No. 12150610462).
{The input and output files of the $ab$ $initio$ and the mVMC calculations are available at the repository~\cite{data}}
\end{acknowledgements}


\begin{thebibliography}{62}
\bibitem [{DFT()}]{DFT}%
  \BibitemOpen
  \href@noop {} {}\bibinfo {note} {{We use the generalized gradient
  approximation (GGA) is employed to calculate the exchange-correlation
  function. The cutoff energies about the wave functions and charge densities
  are set to 80 and 320 Ry, respectively. The wavenumber $\bm k$ mesh is
  $5\times5\times3$.}}\BibitemShut {Stop}%
\end{thebibliography}

\providecommand{\hyphen}{-}\providecommand{\noopsort}[1]{}

\clearpage
\noindent
{\Large Supplemental Materials for ``Gap opening mechanism for correlated Dirac electrons in organic compounds $\alpha$-(BEDT-TTF)$_2$I$_3$ and $\alpha$-(BEDT-TSeF)$_2$I$_3$''}

\section{S.1.~Details of microscopic parameters in $ab$ $initio$ Hamiltonians}
\begin{figure}[h]
\begin{centering}
\includegraphics[width=85mm]{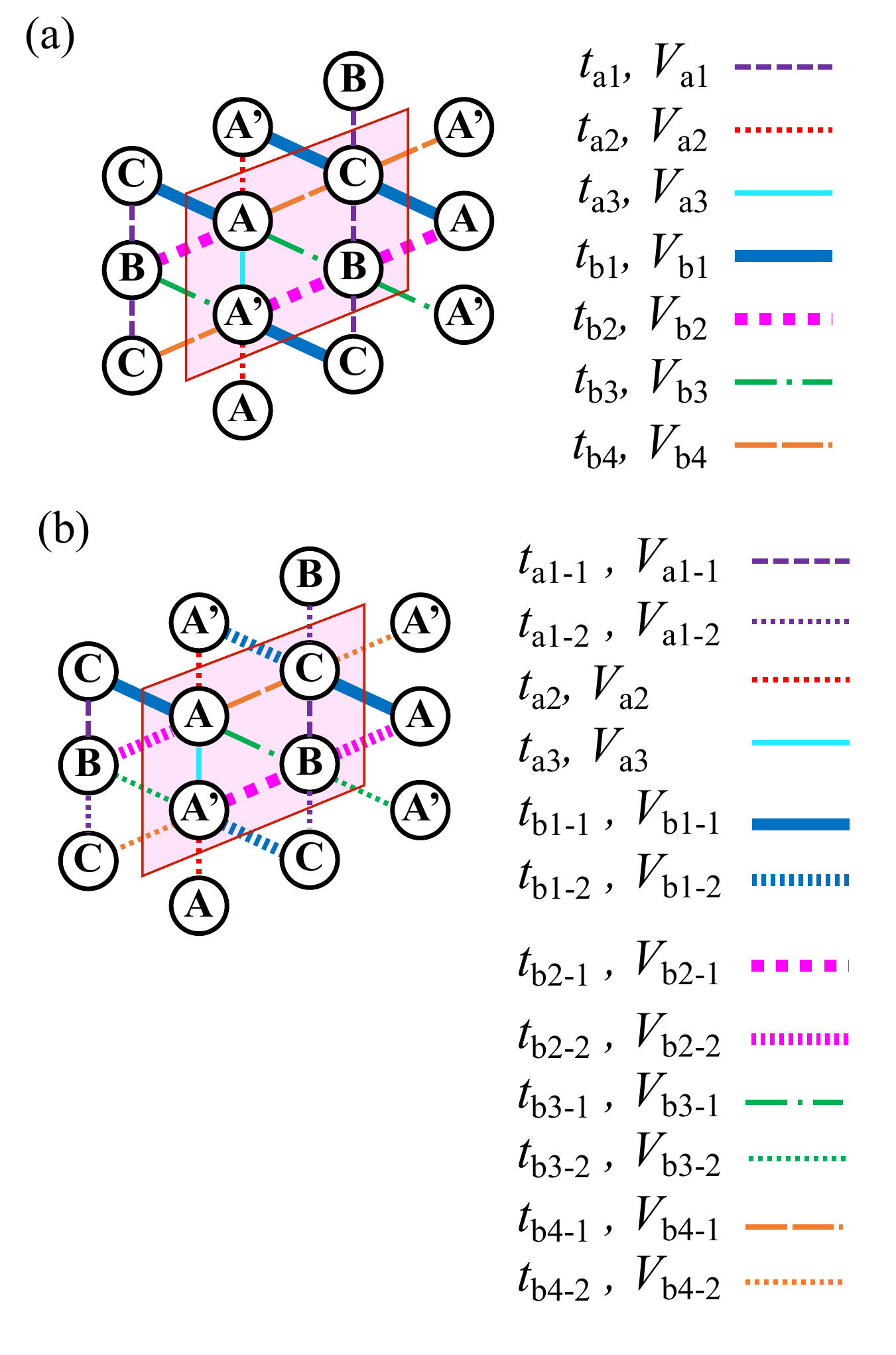}
\caption{(Color online) 
(a) Schematic of lattice structure with inversion symmetry for $\alpha$-(ET)$_2$I$_3$ at 150 K and $\alpha$-(BETS)$_2$I$_3$ at 30 K. (b) Schematic of lattice structure without inversion symmetry for $\alpha$-(ET)$_2$I$_3$ at 30 K. The transfer integrals and Coulomb interactions for the nearest-neighbor sites are also shown. The pink shaded parallelogram shows a unit cell.
}\label{Fig:lattice}
\end{centering}
\end{figure}

Table~\ref{tab:TransWeff} presents the transfer integrals and Coulomb interactions
for $\alpha$-(ET)$_2$I$_3$ at 150 K and $\alpha$-(BETS)$_2$I$_3$ at 30 K, which have inversion symmetry, as shown in Fig. \ref{Fig:lattice} (a). 
Table. \ref{param} presents those for $\alpha$-(BETS)$_2$I$_3$ at 30 K.
For $\alpha$-(ET)$_2$I$_3$ at 30 K, inversion symmetry is broken due to charge ordering. Thus, the relations of the bonds are more complicated, as shown in Fig.\ref{Fig:lattice} (b).

\begin{table*}[h]
\centering
\caption{List of  transfer integrals $t$, difference in chemical potentials $\delta\mu_{\rm iC}\equiv \mu_{\rm i}-\mu_{\rm C}$ (i=A, A$'$, and B), and effective Coulomb interactions $U$ and $V$ (units of meV) for 
 $\alpha$-(ET)$_2$I$_3$ at 150 K and $\alpha$-(BETS)$_2$I$_3$ at 30 K. The values of $t$ and $V$ were averaged between the corresponding bonds shown in Fig. \ref{Fig:lattice}(a).}\label{tab:TransWeff}
\begin{ruledtabular}
  \begin{tabular}{ccc}
   \rule[-3.7mm]{0mm}{8mm}
    {\rm parameters } &
    $\alpha$-(ET)$_2$I$_3$ (150 K) 
    & $\alpha$-(BETS)$_2$I$_3$ (30 K) \\
    \hline \hline
   $t_{\rm a1}$ & $-$17.09 & 9.898 \\ 
   $t_{\rm a2}$ & $-$36.10& $-$16.41 \\
   $t_{\rm a3}$ & 35.58 & 51.07 \\ 
   $t_{\rm b1}$ & 109.5 & 138.2 \\ 
   $t_{\rm b2}$ & 124.7& 158.4 \\ 
   $t_{\rm b3}$ & 46.49 & 65.69 \\ 
   $t_{\rm b4}$ & 14.53 & 18.60 \\ 
   \hline 
   $\delta\mu_{\rm AC}$  & 3.15 &  $-$7.059 \\ 
   $\delta\mu_{\rm A'C}$ & 2.48 &  $-$7.816 \\ 
   $\delta\mu_{\rm BC}$  & 5.28 & $-$13.17 \\ 
\hline 
   $U_{\rm A}$  & 1750 & 1389 \\ 
   $U_{\rm A'}$ & 1750 & 1389 \\ 
   $U_{\rm B}$  & 1772 & 1405 \\ 
   $U_{\rm C}$  & 1738 & 1358 \\
   \hline
   $V_{\rm a1}$ & 662.6 & 580.5 \\
   $V_{\rm a2}$ & 686.9 & 596.2 \\
   $V_{\rm a3}$ & 636.7 & 566.7 \\
   $V_{\rm b1}$ & 636.2 & 579.9 \\
   $V_{\rm b2}$ & 625.7 & 572.9 \\
   $V_{\rm b3}$ & 583.2 & 537.8 \\
   $V_{\rm b4}$ & 608.3 & 556.9 \\
\end{tabular}
\end{ruledtabular}
\label{param}
\end{table*}

\begin{table*}[h]
\centering
\caption{List of transfer integrals $t$, difference im chemical potentials $\delta\mu_{\rm iC}\equiv \mu_{\rm i}-\mu_{\rm C}$ (i=A, A$'$, and B), and effective Coulomb interactions $U$ and $V$ (units of meV) for 
 $\alpha$-(ET)$_2$I$_3$ at 30 K. The corresponding bonds are shown in Fig. \ref{Fig:lattice}(b).}\label{tab:TransWeff2}
\begin{ruledtabular}
  \begin{tabular}{cl}
   \rule[-3.7mm]{0mm}{8mm}
    {\rm parameters } &
    $\alpha$-(ET)$_2$I$_3$ (30 K) \\
    \hline \hline
   $t_{\rm a1-1}, t_{\rm a1-2}$ & $-$30.12, $-$5.250\\ 
   $t_{\rm a2}$ & $-$37.51 \\
   $t_{\rm a3}$ & 38.07 \\ 
   $t_{\rm b1-1},t_{\rm b1-2}$ & 97.48, 124.6 \\ 
   $t_{\rm b2-1},t_{\rm b2-2}$ & 136.2, 125.5 \\ 
   $t_{\rm b3-1}, t_{\rm b3-2}$ & 44.92, 44.96 \\ 
   $t_{\rm b4-1}, t_{\rm b4-2}$ & 24.17, 1.806 \\ 
   \hline 
   $\delta\mu_{\rm AC}$  & 1.111  \\ 
   $\delta\mu_{\rm A'C}$ & 12.87  \\ 
   $\delta\mu_{\rm BC}$  & 22.44  \\ 
\hline 
   $U_{\rm A}$  & 1735 \\ 
   $U_{\rm A'}$ & 1732 \\ 
   $U_{\rm B}$  & 1763 \\ 
   $U_{\rm C}$  & 1722 \\
   \hline
   $V_{\rm a1-1}, V_{\rm a1-2}$ & 650.1, 652.9 \\
   $V_{\rm a2}$ & 676.5 \\
   $V_{\rm a3}$ & 626.6 \\
   $V_{\rm b1-1}, V_{\rm b1-2}$ & 612.8, 634.4\\
   $V_{\rm b2-1}, V_{\rm b2-2}$ & 618.5, 613.1 \\
   $V_{\rm b3-1}, V_{\rm b3-2}$ & 577.5, 564.2 \\
   $V_{\rm b4-1},V_{\rm b4-2}$ & 606.6, 591.4 \\
\end{tabular}
\end{ruledtabular}
\label{param}
\end{table*}

\clearpage

\section{S.2.~Stability of charge-ordered state in $\alpha$-(ET)$_2$I$_3$ at 150K}
\begin{figure}[h]
\begin{centering}
\includegraphics[width=85mm]{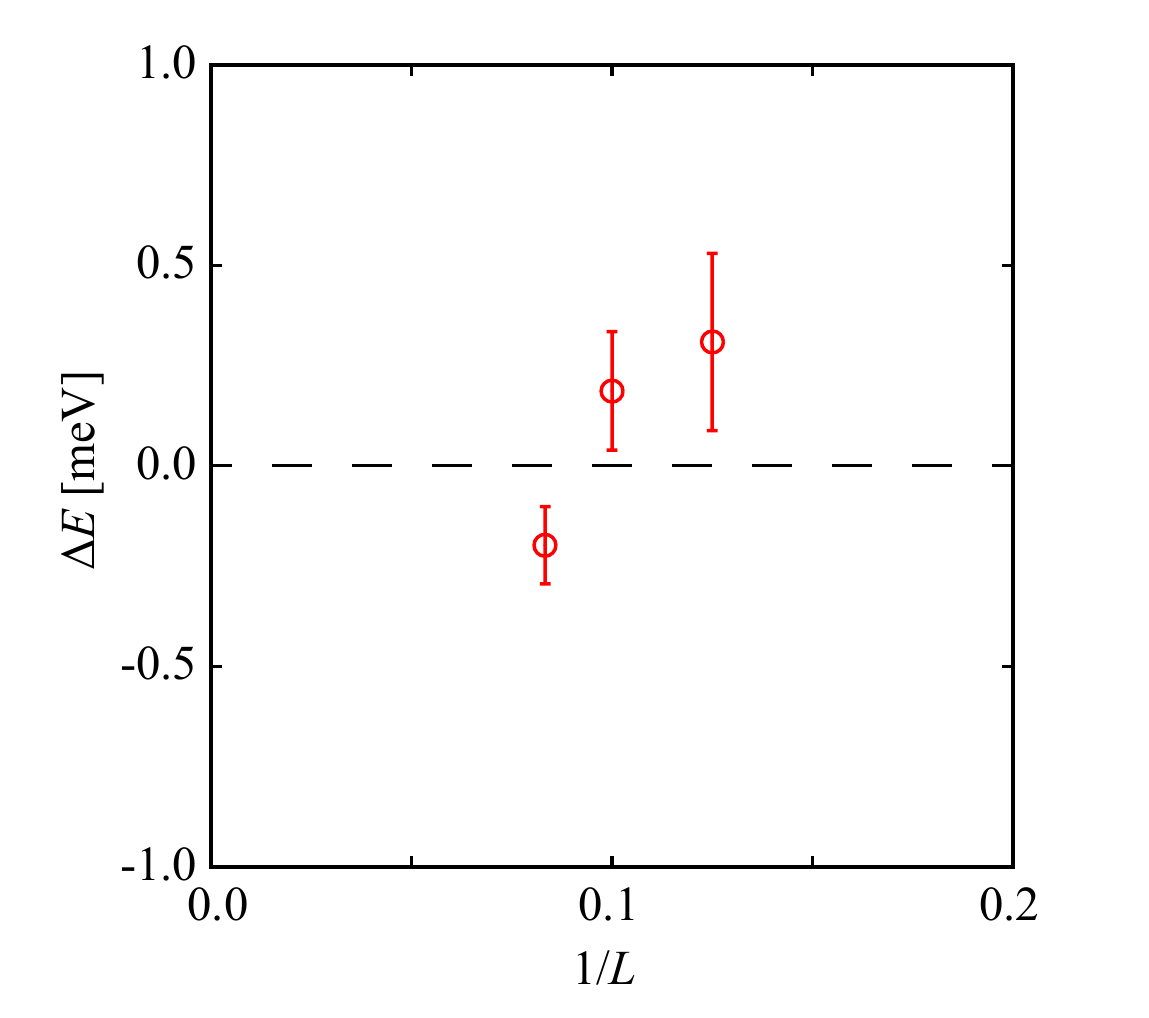}
\caption{(Color online) 
Energy difference between HCO and non-HCO states defined
as $\Delta E= E_{\rm HCO}/N_{\rm s}-E_{\rm non-HCT}/N_{\rm s}$.
}\label{Fig:DEne}
\end{centering}
\end{figure}

We solve the effective Hamiltonians for $\alpha$-(ET)$_2$I$_3$ with 150 K structures
using mVMC. The forms of the wavefunctions for mVMC are the same as those explained in
the main text. In the Hamiltonian for the 150 K structure, since the inversion symmetry
is not broken, the chemical potentials for the A and A$^{\prime}$ sites are equivalent
within the order of meV. For this Hamiltonian, we obtain two different states, the horizontal stripe charge ordered (HCO) state and non-HCO state for $L\geq8$. 
For $L=12$, the charge densities for the HCO state are given by
$\langle n_{\rm A}\rangle=1.47$,
$\langle n_{\rm A'}\rangle=1.49$,
$\langle n_{\rm B}\rangle=1.52$,
and $\langle n_{\rm C}\rangle=1.51$ and
the charge densities for the non-HCO state are
$\langle n_{\rm A}\rangle=1.48$,
$\langle n_{\rm A'}\rangle=1.48$,
$\langle n_{\rm B}\rangle=1.52$,
and $\langle n_{\rm C}\rangle=1.51$.
The statistical errors in the Monte Carlo sampling for the electron densities are of the order of $10^{-3}$.
We confirm that the size dependence of the charge densities is negligibly small.
The difference in the charge densities for the A and A$^{\prime}$ sites is about 0.02, and is smaller than that obtained for the Hamiltonians with a 30 K structure ($\sim0.06$).
We also find that the HCO and non-HCO states are almost degenerate and their energy 
difference is below 1 meV, as shown in Fig.~\ref{Fig:DEne}. 
This result indicates that the lattice distortion
also plays an important role in stabilizing the HCO states although
the Coulomb interactions alone induce instability toward the HCO state.

\end{document}